\documentclass[aps,twocolumn,prl,showpacs]{revtex4}

\usepackage{graphicx}
\usepackage{amsmath}

\begin{document}

\title{Role of strong correlation in the recent ARPES experiments for 
cuprate superconductors}

\author{S.~Yunoki$^1$, E. Dagotto$^{2,3}$, and S.~Sorella$^1$,}

\affiliation{
$^1$Istituto Nazionale per la Fisica della Materia and International 
School for Advanced Studies, via Beirut 4, 34014 Trieste, Italy\\
$^2$Department of Physics and Astronomy, The University of Tennessee, 
Knoxville, Tennessee 37996-1200, USA\\
$^3$Condensed Matter Sciences Division, Oak Ridge National Laboratory, 
Oak Ridge, Tennessee 37831-6393, USA
}

\date{\today}

\begin{abstract}

Motivated by recent photoemission experiments on cuprates, the low-lying 
excitations of a strongly correlated superconducting state are studied 
numerically. It is observed that along the nodal direction these
low-lying one-particle  
excitations show a linear momentum dependence for 
a wide range of excitation energies and, thus, they do not present a kink-like structure. 
The nodal Fermi velocity $v_{\rm F}$, as well as other observables, 
are systematically evaluated directly from the calculated dispersions, and 
they are found to compare well with experiments. 
It is argued that the parameter dependence of $v_{\rm F}$ is 
quantitatively explained by a simple picture of a renormalized Fermi 
velocity. 
\end{abstract}


\maketitle


Since the discovery of the copper-based high-$T_{\rm c}$ 
superconductors~\cite{bed}, there have been extensive 
studies, both experimentally and theoretically, to understand the origin of 
the superconductivity as well as the unusual normal state 
properties~\cite{elbio}. This vast effort has proved
that among the many experimental techniques angle-resolved 
photoemission spectroscopy (ARPES) is one of the most powerful tools, since
it can provide important microscopic information of the electronic 
structure of these materials~\cite{arpes}. One of the recent key findings by ARPES experiments 
concerns the low-lying electronic excitations in the $(0,0)$ to 
$(\pi,\pi)$ direction (nodal direction) of the 
Brillouin zone: (i) the low-lying dispersion shows a kink at 
an energy in the range 50--80 meV from the Fermi 
level~\cite{kink,lanzara}, 
and (ii) the 
nodal Fermi velocity shows almost no doping dependence within the 
experimental error bars ($\sim 1.8\pm0.4$ eV$\cdot$\AA ) for a wide range of 
hole concentrations $x$ ($0<x\alt0.2$--0.3)~\cite{zhou}. 
To explain the peculiar feature (i), the idea that electrons could be strongly coupled to 
other degrees of freedom, such as phonons and magnetic fluctuations, has been 
introduced~\cite{lanzara}. This can naturally explain the appearance of a
new energy scale. There are already theoretical studies 
in this direction~\cite{theory}. However, the  kink structure origin 
is still under debate. 
Regarding feature (ii), there have been no consensus on its origin. 
We believe that before studying more complicated models it is important 
to understand to what extend a purely electronic model alone can 
explain these features observed experimentally. This is precisely the main 
purpose of this work. 

In this paper, using a variational Monte Carlo (MC) method, the low-lying 
excitations of a strongly correlated superconducting state with $d$-wave 
pairing symmetry are studied. By directly calculating the excitation spectrum, 
it is found that the low energy one-particle excitations in the nodal 
direction show a linear momentum dependence, instead of 
a kink-like structure. Our detailed and systematic calculations for the 
nodal Fermi velocity $v_{\rm F}$, as well as a coupling strength 
$\lambda$ defined below, reveals that in spite of not having 
a kink-structure in 
$\varepsilon({\bf k})$ nevertheless the doping dependence of $v_{\rm F}$ and $\lambda$ 
are in good agreement with experiments. Moreover, it is shown that the doping 
dependence of $v_{\rm F}$ can be understood quantitatively as a 
renormalized Fermi velocity.

As a canonical model for the cuprates, here we consider the two dimensional (2D) 
$t$-$J$ model on a square lattice described by the following 
Hamiltonian~\cite{rice}; 
\begin{eqnarray}
H &=& J \sum_{ \langle i,j \rangle } \left( {\bf S}_i \cdot {\bf S}_j 
   -n_in_j/4 \right)
   - t \sum_{ \langle i,j \rangle \sigma } \left( c^{\dag}_{i,\sigma} 
              c_{j,\sigma} + {\rm H.c.}  \right ) \nonumber \\
  &&- t' \sum_{ \langle i,j \rangle' \sigma } 
  \left( c^{\dag}_{i,\sigma} 
              c_{j,\sigma} + {\rm H.c.}  \right ).
\label{tj}
\end{eqnarray}
Here $c^{\dag}_{i,\sigma}$ is the creation operator of a spin 
$\sigma(=\uparrow,\downarrow)$ electron at site $i$, and 
$n_i=n_{i,\uparrow}+n_{i,\downarrow}$ and 
${\bf S}_i={\frac{1}{2}}c^{\dag}_{i,\alpha}{\bf \sigma}_{\alpha\beta} 
c_{i,\beta}$ are the electron density and spin operators. 
$\langle i,j \rangle$ ($\langle i,j \rangle'$) runs over the (next) 
nearest-neighboring sites, and no double occupancy is allowed on each site. 
This model has been studied extensively and found to show a $d$-wave superconducting 
regime in its phase diagram~\cite{sandro,super}.

It is well known that a Gutzwiller projected BCS wave function 
with $d$-wave singlet pairing provides 
a satisfactory variational state for the 2D $t$-$J$ model over a 
wide range of parameters~\cite{gros}. Here we use a slightly more 
complex variational wave function~\cite{sandro} defined by 
$|{\bf\Psi}_{\rm var}^{(N)}\rangle={\hat{\cal P}}_N{\hat{\cal P}}_G
{\hat{\cal P}}_J|{\rm BCS}\rangle$, 
where $|{\rm BCS}\rangle$ is the BCS ground state wave function, 
$|{\rm BCS}\rangle=
\prod_{\bf k}\left[1+f_{\bf k}c_{{\bf k}\uparrow}^\dagger
c_{-{\bf k}\downarrow}^\dagger\right]|0\rangle$, 
$c_{{\bf k},\sigma}^\dagger$ the Fourier transform of $c^{\dag}_{i,\sigma}$, 
${\hat{\cal P}}_N$ the projection operator onto the subspace of $N$ 
electrons, ${\hat{\cal P}}_G=\prod_i(1-n_{i,\uparrow}n_{i,\downarrow})$ 
the Gutzwiller projection operator,  
${\hat{\cal P}}_J=
\exp\left(\sum_{i,j}\alpha^{ij}_{\rm var}n_in_j\right)$ 
a Jastrow factor, and 
$f_{\bf k}=\Delta_{\bf k}/\left(\xi_{\bf k}-{\cal E}_{\bf k}\right)$
with $\Delta_{\bf k}=\Delta_{\rm var}(\cos{k_x}-\cos{k_y})$, 
$\xi_{\bf k}=\varepsilon_{\bf k}-\mu_{\rm var}$, 
$\varepsilon_{\bf k}=-2(\cos{k_x}+\cos{k_y})
-4t_{\rm var}'\cos{k_x}\cos{k_y}$, and 
${\cal E}_{\bf k}=-\sqrt{\xi_{\bf k}^2+\Delta_{\bf k}^2}$~\cite{shiba,note2}. 
The variational parameters, all the independent pairs of 
$\alpha^{ij}_{\rm var}$, $\Delta_{\rm var}$, $\mu_{\rm var}$, and 
$t_{\rm var}'$, are determined by minimizing the variational energy: 
$E({\bf\Psi}_{\rm var}^{(N)})
=\langle{\bf\Psi}_{\rm var}^{(N)}|H|{\bf\Psi}_{\rm var}^{(N)}\rangle
/\langle{\bf\Psi}_{\rm var}^{(N)}|{\bf\Psi}_{\rm var}^{(N)}\rangle$ 
for $N$ even~\cite{sorella}. In this case $|{\bf\Psi}^{(N)}_{\rm var}\rangle$ 
is a spin singlet and has a well-defined total momentum zero. 
Hereafter, the wave function with the optimized parameters is denoted by 
$|{\bf\Psi}^{(N)}\rangle$, and $E^{(N)}=E({\bf\Psi}^{(N)})$. 
Also the energy unit ($t$) and the lattice constant $a$ are both set to 
be one.

A single hole added to $|{\bf\Psi}^{(N)}\rangle$ is naturally described  
by 
$|{\bf\Psi}^{(N-1)}_{\bf k}\rangle={\hat{\cal P}}_{N-1}{\hat{\cal P}}_G
{\hat{\cal P}}_J \gamma_{{\bf k}\uparrow}^\dagger |{\rm BCS}\rangle$, 
where $\gamma_{{\bf k}\uparrow}^\dagger$ 
is the creation operator of the standard Bogoliubov quasiparticle with 
momentum ${\bf k}$ and $s$=$\uparrow$~\cite{text}. 
The state with a single added electron,
$|{\bf\Psi}^{(N+1)}_{\bf k}\rangle$, is described in a similar
way replacing ${\hat{\cal P}}_{N-1}$ 
by ${\hat{\cal P}}_{N+1}$. Note that these states have sharply defined 
${\bf k}$, total spin 1/2, and $z$-component of total spin 1/2. The 
use of these states is partially motivated by 
the diagonalization of $H$ on small clusters -- effort that  indicated that the low-lying 
single-particle excitations are well described by a renormalized 
Bogoliubov quasiparticle state~\cite{eder} -- and partially on the recent 
proposal of a similar state~\cite{bob}. The variational 
energies for these $(N\pm1)$-electron states are denoted by  
$E^{(N\pm1)}_{\bf k}$. The  
single-particle excited states dispersion is thereby evaluated using 
$\varepsilon({\bf k})$=$E^{(N)}-E^{(N-1)}_{\bf k}$ 
($E^{(N+1)}_{\bf k}-E^{(N)}$) for the one-electron removal (addition) 
spectrum~\cite{elbio}.

\begin{figure}[hbt]
\includegraphics[width=5.7cm,angle=-90]{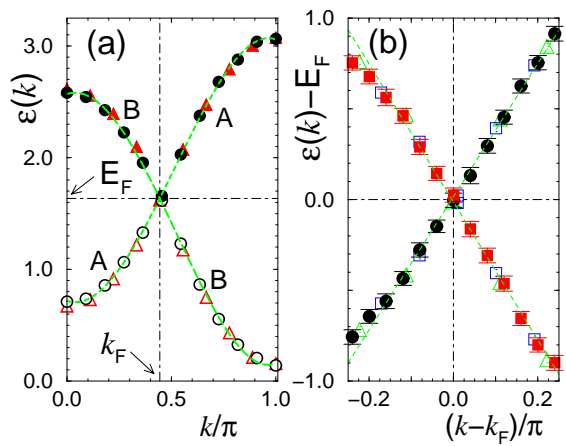}
\begin{center}
\caption{
One-particle dispersion $\varepsilon({\bf k})$ in the nodal direction, 
${\bf k}=(k,k)$, for the 2D $t$-$J$ model with $J/t$=$0.3$ and $t'/t$=$-0.2$, at 
$x$=$0.099$. (a): Full dispersion for $L$=$162$ (triangles) and $242$ 
(circles). The one-electron removal (addition) spectrum is denoted 
by open (solid) symbols. The dashed (long dashed) line is a fitting curve 
of branch A (B) for the $L$=$242$ data, using up to 3rd order polynomials. 
The estimated Fermi momentum and energy are indicated by 
$k_{\rm F}$ and 
$E_{\rm F}$, respectively. (b): Same as (a) but focusing on the 
excitations near $E_{\rm F}$. 
In addition to the data for $L$=$162$ (open triangles) and $242$ (open squares), 
results for $L=1250$ (solid squares and circles) are also plotted. Dotted 
lines are a guide to the eye. 
}
\label{disp}
\end{center}
\end{figure}

In Fig.~\ref{disp} (a), as a typical example, the calculated dispersion 
$\varepsilon({\bf k})$ in the nodal direction is shown for the 2D 
$t$-$J$ model with $J/t=0.3$ and $t'/t=-0.2$, at hole density 
$x$=$1-n$=$0.099$.
Here $n$=$N/L$ and $L$ is the total size of the system. 
In the figure, the one-particle removal and additional spectra are denoted by 
open and solid marks, respectively. As seen in the figure, these spectra are 
almost symmetric in energy about the center of the dispersion. 
This might be expected because $|{\bf\Psi}^{(N\pm1)}\rangle$ is made of a 
single Bogoliubov mode (with the projections) which shows a symmetric spectrum 
$\pm{\cal E}_{\bf k}$~\cite{text}. 
From these results several important quantities are evaluated, 
such as the nodal Fermi momentum (${\bf{k}}_{\rm F}=(k_{\rm F},k_{\rm F})$), 
Fermi energy ($E_{\rm F}$), bandwidth ($W$), as well as the nodal Fermi 
velocity at ${\bf k}_{\rm F}$ ($v_{\rm F}$).

Let us refer to branch A as the main dispersion and to branch B as the
''shadow'' 
dispersion, since branch A (branch B) consists of all the one-electron removal 
(additional) states inside $k_{\rm F}^{\rm var}$, and all the one-electron 
additional (removal) states outside $k_{\rm F}^{\rm var}$. 
Here, $k_{\rm F}^{\rm var}$ is the nodal Fermi point of $|{\rm BCS}\rangle$ 
with the optimized parameters. Although this assignment is quite natural, 
we also calculated the quasi-particle weight directly and found that branch A has 
substantially more weight than branch B. 
The energy 
difference between ${\bf k}=(0,0)$ and ${\bf k}=(\pi,\pi)$ in each branch 
naturally defines the bandwidth $W=\varepsilon(\pi,\pi)-\varepsilon(0,0)$. 
Next, we fit the data in each branch using up to 
third-order polynomials. As shown in Fig.~\ref{disp}(a), the fitting is highly 
satisfactory. The intersection of the fitting curves provides the Fermi 
energy $E_{\rm F}$ and the nodal Fermi point ${\bf k}_{\rm F}$~\cite{note4}. 
From these fitting curves, we obtain the nodal Fermi velocity $v_{\rm F}$ 
at ${\bf k}_{\rm F}$.

The satisfactory fitting of the dispersions $\varepsilon({\bf k})$ in the 
nodal direction already indicates that $\varepsilon({\bf k})$ 
is a smooth function of ${\bf k}$ and, therefore, it suggests that 
the state used here does not have the kink structure observed experimentally. 
To study this in more detail, we calculated the 
nodal dispersion $\varepsilon({\bf k})$ on a cluster with $L$=$1250$, where the number of  
allowed ${\bf k}$-points in the nodal direction is 50 and, thus, the momentum 
resolution $\delta |{\bf k}|$ is about $0.18/a$~\cite{note3}. 
The results are presented in Fig.~\ref{disp}(b). 
It is fairly clear that the 
dispersions in both branches are almost linear around $E_{\rm F}$, and no 
kink-like structure is seen. If there were a kink structure in the dispersion, 
as observed experimentally, it would not be possible to fit the 
data for both one-electron removal and additional spectra 
using the same straight line. 
Comparing the data for $L=1250$ and those for smaller systems shown in 
Fig.~\ref{disp}(b), it is apparent that the size dependence of the dispersion 
is small, and therefore we can safely estimate quantities such as $v_{\rm F}$ 
using the smaller systems. 

\begin{figure}[hbt]
\includegraphics[width=5.7cm,angle=-90]{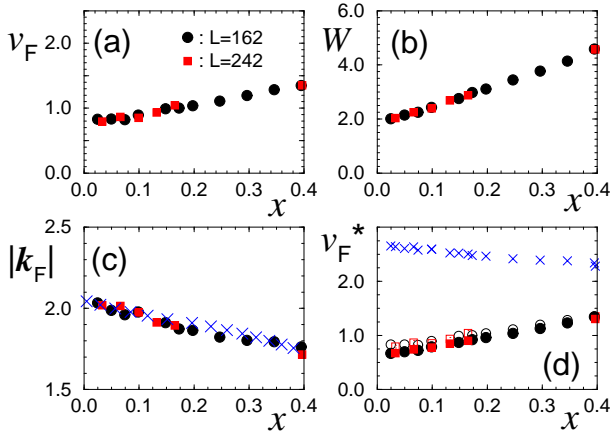}
\begin{center}
\caption{
(a) Nodal Fermi velocity $v_{\rm F}$, (b) bandwidth $W$, (c) nodal 
Fermi momentum $|{\bf k}_{\rm F}|$ (solid symbols), and (d) renormalized Fermi 
velocity $v_{\rm F}^*$ (solid symbols) (see text) for the 2D $t$-$J$ 
model with $J/t$=$0.3$ and $t'/t$=$-0.2$ at different $x$. 
For comparison, in (c) the free-electrons nodal Fermi points 
(crosses) are also shown. In (d) the Fermi velocity for the free 
electrons $v_{\rm F}^0$ (crosses) and $v_{\rm F}$ (open symbols) 
are also plotted. 
}
\label{vf}
\end{center}
\end{figure}

 Figure~\ref{vf} summarizes the $x$ dependence of various quantities for the 2D 
$t$-$J$ model with $J/t$=$0.3$ and $t'/t$=$-0.2$, which 
is a typical parameter set for the cuprates~\cite{elbio}. 
Fig.~\ref{vf} (a) shows that $v_{\rm F}$ is weakly dependent on $x$ up to 
about $x$=0.1--0.2 -- with $v_{\rm F}\sim$ 0.8--1.0 $t$ -- and then 
increases with further increasing $x$. If $t\sim500$ meV and $a\sim4$ {\AA} are 
used, the calculated value of this nearly $x$-independent $v_{\rm F}$ 
corresponds to approximately 1.6--2.0 eV$\cdot${\AA}, which is compatible 
with ARPES data within the experimental error bars~\cite{zhou}. 
The present results are also consistent with recent reports by 
Paramekanti, {\it et al.}~\cite{rand,rand2}, including the overall 
increasing behavior of $v_{\rm F}$ as a function of $x$~\cite{note1}. 

In Fig.~\ref{vf} (b) and (c), the calculated bandwidth $W$ and the nodal 
Fermi momentum $k_{\rm F}$ are shown, respectively. Although 
$k_{\rm F}$ approximately follows the result for free electrons (with $t'$), 
$W$ has a 
stronger $x$ dependence. It should be emphasized that this non-trivial $x$
dependence of $W$ is caused by strong correlations, which are imposed in 
the wave function by the Gutzwiller 
projection ${\hat{\cal P}}_G$. From Fig,~\ref{vf} (b), 
it is expected that the effective mass in the nodal direction monotonically 
increases with decreasing $x$, but it is finite even at $x\to0$.

Now we show that the $x$ dependence of $v_{\rm F}$ is understood 
quantitatively by a simple picture of a renormalized Fermi velocity. 
Since $k_{\rm F}$ is similar to the free electrons results (with $t'$), 
a natural procedure to follow is to calculate a renormalized Fermi 
velocity $v_{\rm F}^*$ from the value of the free electrons 
$v_{\rm F}^0$ at $k_{\rm F}$. 
In Fig.~\ref{vf} (d), $v_{\rm F}^*=\gamma v_{\rm F}^0$ is 
plotted, where $\gamma=W/W_0$ is a renormalization constant and 
$W_0$=$8t$ is the free electrons bandwidth. 
Clearly, $v_{\rm F}^*$ can now reproduce $v_{\rm F}$ for almost 
all the doping range studied.

\begin{figure}[hbt]
\includegraphics[width=7.5cm,angle=0]{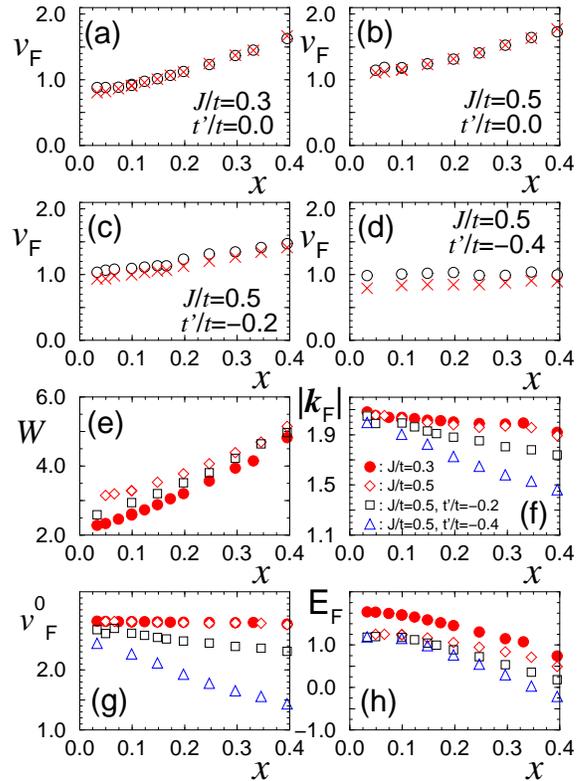}
\begin{center}
\caption{
(a)--(d): $x$ dependence of the nodal Fermi velocity $v_{\rm F}$ 
(open marks) and the renormalized Fermi velocity $v_{\rm F}^*$ (crosses) 
for the 2D 
$t$-$J$ model. The parameters used are indicated in the figures. 
(e)--(h): $x$ dependence of $W$ (e), $k_{\rm F}$ (f), the nodal Fermi 
velocity of the free electrons at ${\bf k}_{\rm F}$ (g), and $E_{\rm  F}$ (h) 
for the different parameters indicated in (f). 
}
\label{vfall}
\end{center}
\end{figure}

To support this argument, systematic calculations are done for various 
model parameters, and the results are shown in Fig.~\ref{vfall} (a)--(d). 
It is apparent from the figures that $v_{\rm F}^*$ can indeed explain 
the $x$ dependence of $v_{\rm F}$ quantitatively for a wide hole-doping range. 
The main features of the $J$ and $t'$ dependences are as follows: 
(i) $v_{\rm F}$ increases with $J$ [Fig.~\ref{vfall} (a) and (b)], and 
(ii) the increasing tendency of $v_{\rm F}$ with $x$ weakens with $|t'|$ 
[Fig.~\ref{vfall} (b)--(d)]. These dependences can be explained by 
the renormalized velocity picture as well. While $k_{\rm F}$ does not depend on 
$J$ [Fig.~\ref{vfall}(f)] and, thus, neither does $v_{\rm F}^0$ 
[Fig.~\ref{vfall}(g)], $W$ does depend on $J$ and becomes 
larger with $J$ [Fig.~\ref{vfall}(e)]. Therefore $v_{\rm F}^*$ 
increases with increasing $J$. In contrast, $t'$ does affect the value of 
$v_{\rm F}^0$, and has a decreasing trend with increasing $x$ 
[Fig.~\ref{vfall}(g)]. Thus, 
$v_{\rm F}^*$ shows a reduced tendency to increase with $x$. Moreover, this decreasing 
trend can cancel the increasing behavior of $W$ with $x$ and, as a consequence,
$v_{\rm F}^*$ 
can present a rather weak $x$ dependence over a wide hole-doping range, as in 
Fig.~\ref{vfall}(d), which agrees qualitatively with the calculated 
dependence of $v_{\rm F}$. 
For comparison with experiments, we also show $E_{\rm F}$ for various 
$t'$ in 
Fig.~\ref{vfall} (h). The $x$ dependence of $E_{\rm F}$ seems to be 
stronger for large $|t'|$ than for small $|t'|$, a trend that has been
seen in experiments~\cite{fujimori}.

\begin{figure}[hbt]
\includegraphics[width=3.2cm,angle=-90]{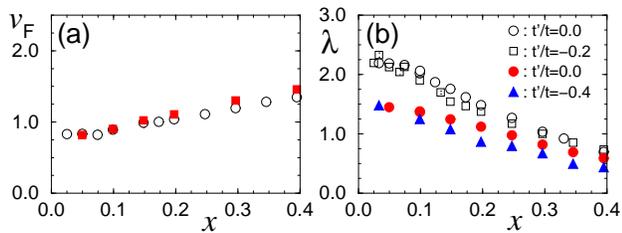}
\begin{center}
\caption{
(a) $v_{\rm F}$ vs. $x$ 
for the 2D $t$-$J$ model with $J/t$=$0.3$ and $t'/t$=$-0.2$ calculated using the Green 
function MC method with fixed node approximation (solid symbols). 
For comparison, the variational estimates shown in Fig.~\ref{vf} (a) are also 
provided (open symbols). 
(b) $x$ dependence of $\lambda$ for $J/t$=$0.3$ (open marks) and 
0.5 (solid marks) with different $t'/t$.
}
\label{vfv}
\end{center}
\end{figure}

Even though our main goal is to study the low-lying excitations of 
the strongly correlated superconducting state, 
$|{\bf\Psi}^{(N\pm1)}_{\bf k}\rangle$, it is interesting to consider the 
accuracy of our estimate of $v_{\rm F}$ for the 2D $t$-$J$ model. 
To this end, we have carried out a fixed-node approximation Green function MC 
simulation using $|{\bf\Psi}^{(N\pm1)}_{\bf k}\rangle$ as guiding 
function~\cite{sorella}. As observed in Fig.~\ref{vfv} (a), the calculated 
$v_{\rm F}$ compare well with the variational estimate~\cite{var}.

Finally, we show in Fig.~\ref{vfv} (b) a coupling strength defined as
$\lambda$=$v_{\rm F}^0/v_{\rm F}-1$ which is the first derivative of 
the real part of the single-hole self energy $\Sigma'$ at $E_{\rm F}$ 
($\lambda$=$-\partial\Sigma'(\omega)/\partial\omega|_{\omega=E_{\rm F}}$) 
if the momentum 
dependence of $\Sigma'$ is assumed weak~\cite{mahan,fink,sigma}. 
It is interesting to note that $\lambda$ seems to be determined 
largely by $J$, not by $t'$, for a wide range of $x$, and 
becomes weakly depending on $J$ and $t'$ for $x\agt0.3$--0.35. 
This monotonically decreasing behavior with $x$ as well as the value 
of $\lambda$ are in good 
agreement with experimental estimates~\cite{lanzara}.

In conclusion, the low-lying one-particle excitations 
of a strongly correlated superconducting state was studied. 
It was found that the dispersion $\varepsilon({\bf k})$ in the nodal 
direction shows a
linear dependence for a wide range of excitation energies around 
$E_{\rm F}$ ($\sim$1.0--1.5$t$) and, thus, does  not present 
a kink-like structure~\cite{kink2}. 
Systematic estimations are made for the nodal Fermi 
velocity $v_{\rm F}$ directly from the calculated $\varepsilon({\bf k})$. 
The $x$ dependence of $v_{\rm F}$ as well as $\lambda$ are 
in good agreement  with experiments. It is shown that the model 
parameter dependence of $v_{\rm F}$ is quantitatively explained by 
a simple picture of the renormalized Fermi velocity.  
Our results suggest that the kink structure observed 
experimentally is caused by other degrees of freedom not included 
in our study. Although the calculated $\varepsilon({\bf k})$ in the nodal 
direction has no kink, the estimated 
$v_{\rm F}$ as well as the coupling strength $\lambda$  compare well 
with ARPES experiments. These results lead us to the speculation that 
a major part of the low energy physics for the cuprates can still be 
described mainly by a purely electronic $t$-$J$-like model~\cite{gweon}. 
The ``universal'' Fermi velocity found in experiments~\cite{zhou} turns 
out to be 
explained here by a rather accidental compensation of two effects: the 
bandwidth $W$ decreases with decreasing doping $x$ due to correlation, 
while the bare Fermi velocity instead increases and further changes with 
$t'/t$.
Therefore this effect might be less significant from the  
theoretical point of view, contrary to what was previously assumed. 
Our results indicate that more accurate experiments should eventually detect a 
''non-universal'' and weakly doping dependent Fermi velocity.


We are grateful to C. Castellani, R. Hlubina, N. Nagaosa, and M. Randeria for 
helpful discussions. This work was 
supported in part by MIUR-COFIN 2003. 
E.D. was supported by NSF grants DMR 0122523 and 0312333.
  
%
%
%



\begin{thebibliography}{99}
\bibitem{note2} To stabilize the numerical calculations, occasionally farther 
neighboring pairing functions are included in $|\Psi^{(N)}_{\rm var}\rangle$.
\bibitem{note4} $k_{\rm F}$ determined in this 
way is compatible with $k_{\rm F}^{\rm var}$.
\bibitem{note3} Since we did not find a significant size dependence of the 
optimized variational parameters, the optimized parameters for $L$=$242$ are 
used for $L$=$1250$.
\bibitem{var} We have also calculated the variance of $H$ for 
$|{\bf\Psi}^{(N\pm1)}_{\bf k}\rangle$ and found that 
the variance is as small as the one for $|{\bf\Psi}^{(N)}\rangle$, 
indicating that $|{\bf\Psi}^{(N\pm1)}_{\bf k}\rangle$ is as good as 
 $|{\bf\Psi}^{(N)}\rangle$.
\bibitem{sigma} By computing the quasi-particle weight, we have checked 
that this widely used assumption holds for $x\agt0.1$.
\end{thebibliography}
\end{document}